\documentclass{article}
\usepackage{spconf,amsmath,graphicx}
\usepackage{multirow}
\usepackage{subfigure}

\title{Advanced Framework for Animal Sound Classification With Features Optimization}
 \twoauthors
 {Qiang Yang, Xiuying Chen, Changsheng Ma, Carlos M. Duarte}
	{King Abdullah University of Science and Technology\\
	Jeddah, Saudi Arabia
	}
 {Xiangliang Zhang}
	{University of Notre Dame\\
	Indiana, United States}
\begin{document}
%
\maketitle
\begin{abstract}

The automatic classification of animal sounds presents an enduring challenge in bioacoustics, owing to the diverse statistical properties of sound signals, variations in recording equipment, and prevalent low Signal-to-Noise Ratio (SNR) conditions. Deep learning models like Convolutional Neural Networks (CNN) and Long Short-Term Memory (LSTM) have excelled in human speech recognition but have not been effectively tailored to the intricate nature of animal sounds, which exhibit substantial diversity even within the same domain. We propose an automated classification framework applicable to general animal sound classification. Our approach first optimizes audio features from Mel-frequency cepstral coefficients (MFCC), including feature rearrangement and feature reduction. It then uses the optimized features for the deep learning model, i.e., an attention-based Bidirectional LSTM (Bi-LSTM), to extract deep semantic features for sound classification. We also contribute an animal sound benchmark dataset encompassing oceanic animals and birds\footnote{https://github.com/gitdevqiang/AnimalSound}. Extensive experimentation with real-world datasets demonstrates that our approach consistently outperforms baseline methods by over 25\% in precision, recall, and accuracy, promising advancements in animal sound classification.

\end{abstract}
\section{Introduction}
\label{sec:intro}

The automatic classification of animal sounds represents a longstanding challenge within the realm of bioacoustics, yet it is a crucial endeavor with far-reaching implications for monitoring biodiversity and ecosystem health through acoustic sensing \cite{gasc2015acoustic,qi2008soundscape,tucker2014linking}. This task is particularly complex due to the inherent variability in sound signals, stemming from diverse statistical properties, the utilization of varying recording equipment, and frequently encountered low SNR conditions.

Previous research in sound classification can be broadly categorized into three groups: traditional machine learning methods \cite{phyu2009survey,fu2010learning,shen2011automatic,wei2020research,mehyadin2021birds,vimal2021mfcc}, deep learning-based approaches \cite{stowell2016bird,rajanna2015deep,dumpala2017improved,fang2022fast,xie2022sliding,tsalera2022cnn}, and image analysis-based techniques \cite{fu2010survey,paraskevas2003audio,amiriparian2017snore,ren2018learning,an2022development,cohen2020challenges}. Traditional methods often rely on MFCC \cite{foote1997content} for sound feature extraction, which offers a representation of the short-term power spectrum of sound through a linear cosine transform of a log power spectrum, employing a nonlinear mel-scale of frequency. Fu et al. investigated two classifiers based on naive Bayes for music classification \cite{fu2010learning}. Vimal et al. separated the MFCC attributes or types and used them to coach machine learning algorithms \cite{vimal2021mfcc}. Rajanna et al. leveraged a two-layer neural network for music genre classification to learn the input-output relationship \cite{rajanna2015deep}. Xie et al. proposed a novel feature set for automatically classifying bird sounds based on an optimized scale-frequency map and CNN \cite{xie2022sliding}. Paraskevas et al. used acoustic images from spectrograms and statistical compression to create a feature matrix for classification \cite{paraskevas2003audio}. Ren et al. proposed to apply pre-trained ImageNet on scalogram images of Phonocardiogram (PCG) for heart sound classification \cite{ren2018learning}.

To fulfill the automated animal sound classification, we are faced with a multifaceted challenge. The specific hurdles encompass: 1) Achieving uniform representation for sound recordings of varying lengths and frequencies without incurring information loss; 2) Proficiently extracting crucial features from MFCC feature matrices; and 3) Designing suitable classifiers to leverage the extracted features optimally.

To address this challenge, we present a novel approach aimed at enhancing the effectiveness of animal sound classification. Firstly, we tackle the issue of variable sound record lengths and frequencies by devising a method that optimally rearranges MFCC feature matrices. Particularly, we propose a rearrangement technique involving flattening and extension of MFCC feature matrices. Furthermore, we introduce a data reduction strategy to mitigate the impact of noisy data where an autoencoder model is used to efficiently reduce feature matrix dimensions, effectively filtering out irrelevant information that hinders the extraction of critical features. To capture essential features for classification, we develop a deep learning approach, employing a Bi-LSTM network with an attention mechanism, enabling differential weighting of features to enhance their significance.

Our contributions are summarized as follows: (1) We introduce a versatile automatic animal sound classification system across diverse domains; (2) We propose an innovative sound classifier that contains a feature rearrangement module for exploring bidirectional sequential relations, a feature reduction module for mitigating the impact of noisy data, and a Bi-LSTM module augmented with an attention mechanism, enabling the extraction of profound features; (3) We contribute an animal sound benchmark dataset containing marine animals and birds; and (4) Extensive experimentation on real-world datasets unequivocally demonstrates the superiority of our proposed method, showcasing remarkable improvements.
\section{Methods}
\label{sec:methods}

In the workflow, we first use MFCC to extract feature matrices from audio data \cite{sahidullah2012design}, denoted as 
$M =
  \left[ {\begin{array}{cccc}
    f_{11}  & \cdots & f_{1N}\\
    \vdots  & \ddots & \vdots\\
    f_{D1}  & \cdots & f_{DN}\\
  \end{array} } \right]$.
The matrix is then creatively reshaped to better capture time sequence relations, considering both past and future time points. To combat noise, we conduct feature reduction, preserving essential features while discarding irrelevant ones. Finally, our deep learning network, featuring Bi-LSTM for sequential context and an attention mechanism for feature weighting, utilizes processed MFCC matrices to excel in the classification task.

\subsection{Feature Matrix Rearrangement Module}

Drawing inspiration from Word2Vec \cite{church2017word2vec}, which considers both previous and future contexts in natural language processing, we aim to leverage full-time sequence information for MFCC features. In Word2Vec, to learn the feature vector of a word like ``fox", its neighboring words (``quick," ``brown", ``jumped", and ``over") within a defined window are considered in the form of probabilities. This contextual information helps understand the semantics of the target word ``fox". Similarly, such contextual information is vital for audio data analysis because each MFCC feature, represented as $f_{ij}$, is interconnected with its nearby previous and future time frames due to the presence of semantic relations among them. Inspired by this, we propose to rearrange MFCC feature matrices by recombining element relations. Specifically, we define a smaller slice dimension as $s \in R^{d \times n'}$, where $n'$ is much smaller than s. We then split the original feature matrix into multiple slices using the following equation:
\begin{equation}
    \tilde{M} = slice(M)=\{s_1,s_2,s_3,…,s_m\}
\end{equation}
where $slice()$ is a function to slice up feature matrix and $m=\lceil {N/N'} \rceil$.
Next, we pad the slice whose time frame is not equal to $N'$ by adding a special value, such as 0 on the left side or right side, so we can have several slices of the same size.
\begin{equation}
    ps = padding(s, s') \in R^{D \times N'}
\end{equation}
After that, we reshape every slice into the fixed form by flattening it, which is similar to the word in the sentence below. 
\begin{equation}
    fs = flattening(ps) \in R^{1 \times (D\times N')}
\end{equation}
Therefore, we have a new slice set $FS= \{{fs}_1,{fs}_2,{fs}_3,…,\\{fs}_m\}$ with the size $1\times (D\times N')$. Finally, we recombine the reshaped slices and get the rearranged new feature matrixes $M' \in R^{m \times (D\times N')}$ with the following equation:
\begin{equation}
    M'=recom(FS, axis) 	
\end{equation}
where $axix$ denotes the direction to recombine the new slice, whose values are $x-axis$ and $y-axis$. By doing so, we can consider not only the next future time points but also previous time points, which helps us gain more semantic information from audio data compared with the original method. 

\begin{figure}[h]
    \centering
    \includegraphics[width=0.35\textwidth]{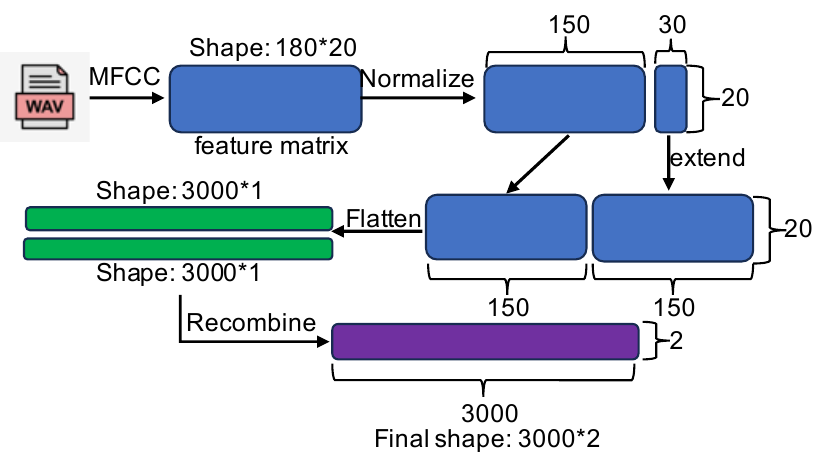}
    \vspace{-0.15in}
    \caption{Toy example of feature matrix rearrangement.}
    \label{fig:toy}
\end{figure}

We provide an example to illustrate the feature matrix rearrangement. We first use MFCC to extract feature matrices from an audio file whose dimension is $180 \ast 20$. Assuming the chosen slice size with the shape $150 \ast 20$, we divide the original feature matrix into two parts: one with dimensions $150 \ast 20$ and the other with $30 \ast 20$. The second slice is extended to match the slice dimension. Then, each slice is flattened to have the dimension $3000 \ast 1$, and finally, they are recombined into a single matrix with dimensions $3000 \ast 2$, as illustrated in Fig. \ref{fig:toy}.

\vspace{-0.2in}

\vspace{-0.1in}
\subsection{Feature Matrix Dimension Reduction Module}
The noisy data significantly impacts classification outcomes, potentially leading to erroneous decisions. In the context of audio data, noise from environmental sources, such as waves, wind, and other species' sounds, is inevitable. Moreover, when dealing with high-dimensional data, noisy data can interfere with and disrupt the attention mechanism. Dimension reduction techniques, such as Principal Component Analysis (PCA) and Linear Discriminant Analysis (LDA) \cite{bouzalmat2014comparative,zhao2009comparative,delac2005independent}, have been extensively explored. However, these methods suffer from data leakage or overlook feature relations. 

Fortunately, the autoencoder, a non-linear method, can capture complex relations, and flexible reduced dimensions, and does not require feature orthogonality. Therefore, we design an autoencoder module comprising an encoder and a decoder to reduce the influence of noisy data. It retains essential features while filtering out irrelevant noisy data, yielding promising results. The encoder and decoder share the same structure implemented by fully connected layers.

\subsection{Bi-LSTM-based Classifier with Attention Module}
To enhance the performance of animal sound classification, we leverage a deep learning model to unearth deep relations and characteristics from data. Given the distinct nature of our new feature matrix, which centers around time sequence, we use the Bi-LSTM network, which excels at handling time sequence data by simultaneously considering information from past (backward states) and future (forward states). Furthermore, it adeptly learns long-term dependencies within time-series data. Additionally, we incorporate the attention mechanism for automatically learning different weights to different data segments. 

Given rearranged and reduced feature matrices, we take them as input for the Bi-LSTM layer to capture the semantic information from two directions. The hidden state $h_j$ is calculated below:
\begin{equation}
\begin{aligned}
    \overrightarrow{h}_j &= \overrightarrow{LSTM}(M'_j)\\
    \overleftarrow{h}_j &= \overleftarrow{LSTM}(M'_j)\\
    h_j &= concat([\overrightarrow{h}_j,\overleftarrow{h}_j])
\end{aligned}
\end{equation}
Next, we design an attention layer to learn the weights of features. Specifically, we have:
\begin{equation}
\begin{aligned}
    c_i = \sum_{j=1}^T\alpha_{ij}h_j\\
    \alpha_{ij} = \frac{exp(e_{ij})}{\sum_{j=1}^{T}exp(e_{ij})}\\
    e_{ij} = tanh(Wh_j+b)
\end{aligned}
\end{equation}
Where W is a weight matrix and b is a bias. Then we use a fully connected layer fed from the attention layer. 
%
Next, we use a fully connected layer followed by a softmax layer to do classification. We utilize a dynamic learning rate and cross-entropy loss to train our model.

\section{EXPERIMENTS}
\label{sec:exp}
\subsection{Experimental Settings}
\textbf{Datasets and Metrics}. Two real-world audio datasets are crawled from two repositories about marine animal audio \footnote{https://cis.whoi.edu/science/B/whalesounds/index.cfm} and bird audio \footnote{https://www.floridamuseum.ufl.edu/bird-sounds/}. In the data pre-processing, we delete the audio files of less than 4 samples for each species and then regularize the names of labels. The final set contains 1233 sample audio data sets of mammals from 32 species and 188 samples from 26 bird species. We randomly separated our two data sets into three parts with a ratio of 7: 2: 1 to be training data, validation data, and testing data, respectively. The number of features for MFCC is set as 20. We utilize precision, recall, and accuracy to measure the performance.
\\
\textbf{Comparative Methods.} To evaluate the effectiveness of our proposed feature rearrangement solution, we apply the same deep learning classifier with the same attention mechanism on the MFCC features without rearrangement. This first baseline is named MD, because it uses single-direction features. Our proposed second baseline method is named MDR, which adds a rearrangement operation to extract double-direction sequential features on the basis of MD. Apart from that, we designed an auto-encoder model to filter out irrelevant information and reduce the influence of noisy data, therefore reducing the data dimension. The improved method we propose is named MDRR on the base of MDR.

\begin{table}[h]
\scriptsize
\caption{Overall Performance on both datasets.}
\begin{tabular}{l|ccc|ccc}
\hline
\multirow{2}{*}{Methods} & \multicolumn{3}{c|}{Marine}                                             & \multicolumn{3}{c}{Bird}                                               \\ \cline{2-7} 
                         & \multicolumn{1}{c|}{Precision} & \multicolumn{1}{c|}{Recall} & Accuracy & \multicolumn{1}{c|}{Precision} & \multicolumn{1}{c|}{Recall} & Accuracy \\ \hline
MD                       & \multicolumn{1}{c|}{0.5701}    & \multicolumn{1}{c|}{0.5501} & 0.5512   & \multicolumn{1}{c|}{0.4705}    & \multicolumn{1}{c|}{0.4000} & 0.4010   \\ \hline
MDR                      & \multicolumn{1}{c|}{0.7991}    & \multicolumn{1}{c|}{0.7975} & 0.7975   & \multicolumn{1}{c|}{0.6700}    & \multicolumn{1}{c|}{0.5601} & 0.5603   \\ \hline
MDRR                     & \multicolumn{1}{c|}{0.8536}    & \multicolumn{1}{c|}{0.8481} & 0.8481   & \multicolumn{1}{c|}{0.7400}    & \multicolumn{1}{c|}{0.6000} & 0.6000   \\ \hline
\end{tabular}
\label{tab:t1}
\end{table}

\vspace{-0.25in}
\subsection{Evaluation Results}
\subsubsection{Overall Evaluation}
In the comprehensive comparison, MDRR, which incorporates data rearrangement based on MD, achieves superior performance in terms of Precision, Recall, and Accuracy, demonstrating an average improvement of approximately 21\%, 20\%, and 20\%, respectively, on both datasets compared to MD shown in Table 1. 
Our MDRR method achieves an average improvement of 5.5\%, 6.0\%, and 6.0\% in Precision, Recall, and Accuracy, respectively compared to MDR. Our results also highlight that the autoencoder model effectively retains vital information while reducing data dimensionality, whereas other models may fail to capture crucial information.

\vspace{-0.1in}
\subsubsection{Parameter Analysis}
\begin{figure}[h]
\centering
\subfigure[Slice length]{
\includegraphics[width=.315\linewidth]{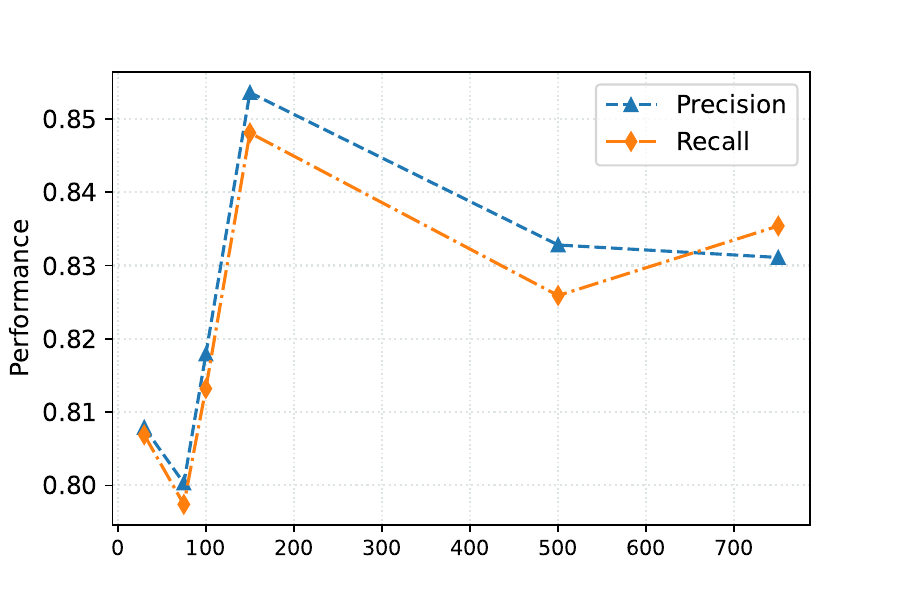}}
\subfigure[Max dimension]{
\includegraphics[width=.315\linewidth]{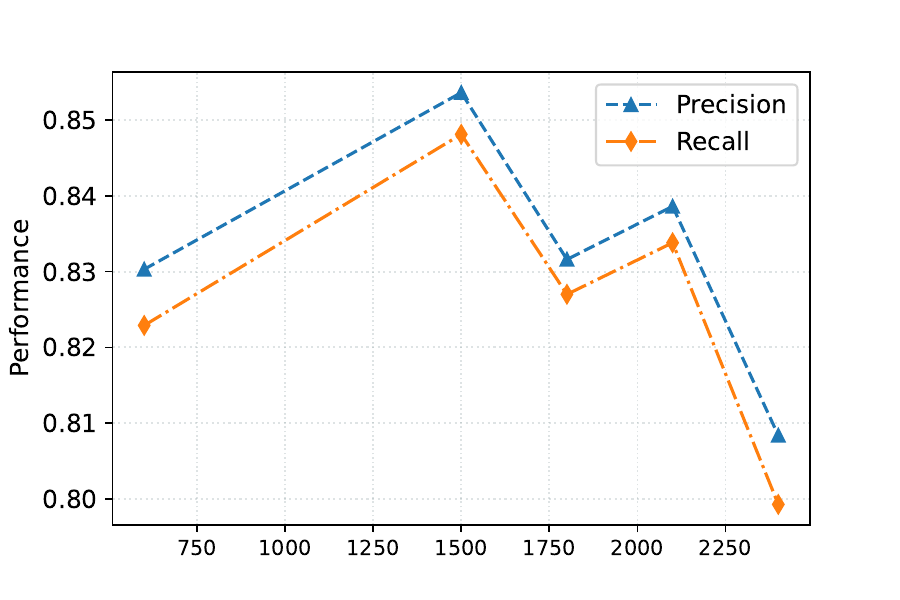}}
\subfigure[Reduced dimension]{
\includegraphics[width=.315\linewidth]{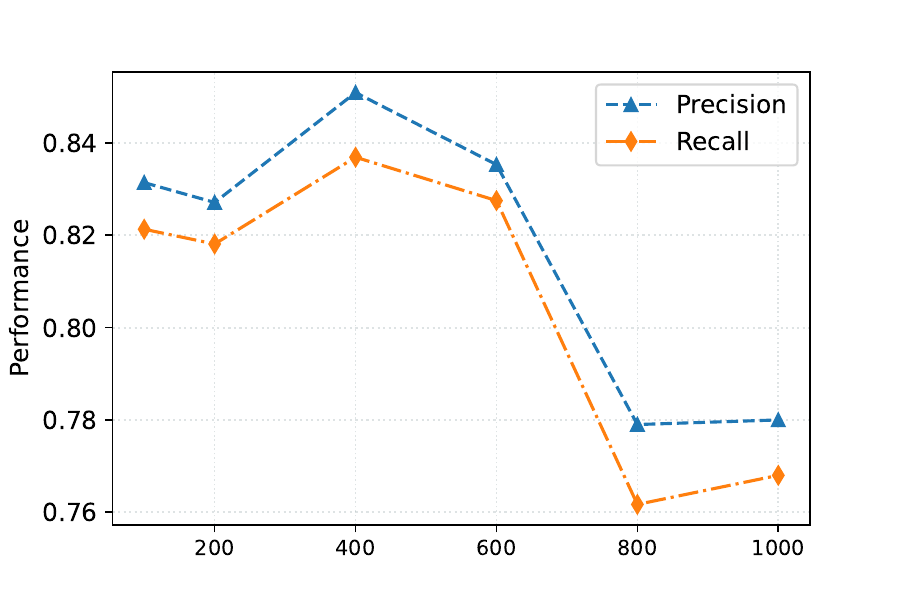}}
\vspace{-0.15in}
\caption{The influence of slice lengths, max dimensions, and reduced dimension of MDRR on the marine animal dataset.}
\label{fig:para}
\end{figure}

We also evaluate the effect of parameters for the performance in  MDRR, including the max dimension of the MFCC feature matrix, slice unit size, autoencoder structure, reduced dimension size, and input shape.
\\
\textbf{Parameters in Data Rearrangement.} 
We maintain a constant reduced dimension of 200 and analyze how the maximum dimension and slice size impact performance. As shown in Fig. \ref{fig:para} (a), when increasing the slice size from 30 to 75, Precision and Recall first have a small decrease. Then they improve greatly reaching 0.854 and 0.848, respectively when increasing to 150. This means that a larger slice size captures more sequentiality of audio data. When it continually grows, the performance has a drop trend due to the involvement of padding information. 
Similarly, when increasing the max dimension from 600 to 2400 shown in Fig. \ref{fig:para} (b), metrics first go up and then decrease. This indicates larger MFCC feature matrix leads to improved ability until an optimum size is reached. Then they drop a lot due to the existence of noise. 
\vspace{-0.1in}
\\
\textbf{Parameter in dimension reduction.} 
Feature dimension reduction significantly reduces the influence of noise. When fixed parameters—slice length at 150, max dimension at 2100, and autoencoder structure at 128 units—were maintained, precision and recall experienced substantial improvements as the reduced dimension increased from 200 to 400, peaking at 0.8509 and 0.8369, respectively shown in Fig. \ref{fig:para} (c). Subsequently, both metrics declined due to the introduction of noise at higher reduced dimensions. 
\\
\textbf{The influence of autoencoder structure.} In our experiments, the paired combinations of slice length and max dimension that performed the best whose value is (150, 1500), especially when the autoencoder structure was set to 128 shown in Fig. \ref{fig:auto}. Overall, structures with output sizes of 128 and 256, utilizing only one dense layer, exhibited superior performance. Deeper layers did not necessarily yield better results and introduced training issues like gradient vanishing and overfitting.
\begin{figure}[h]
\centering
\subfigure[]{
\includegraphics[width=.48\linewidth]{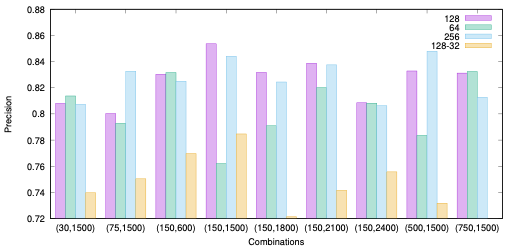}}
\subfigure[]{
\includegraphics[width=.48\linewidth]{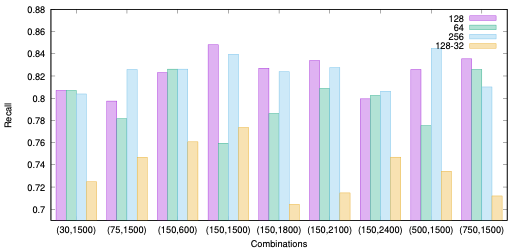}}
\vspace{-0.15in}
\caption{The effect of autoencoder structure of MDRR on the marine animal dataset. 
}
\label{fig:auto}
\end{figure}
\\
\textbf{Effect of the input shape of the classifier.}
The precision, and recall first increase when increasing the input length and decreasing the input\_dim to reach a peak value of 0.8536 and 0.8481 whose combination is (20, 10), and decline at larger values. These results indicate that the input shape of the neural network influences the performance.
\begin{figure}[h]
\centering
\subfigure[]{
\includegraphics[width=.48\linewidth]{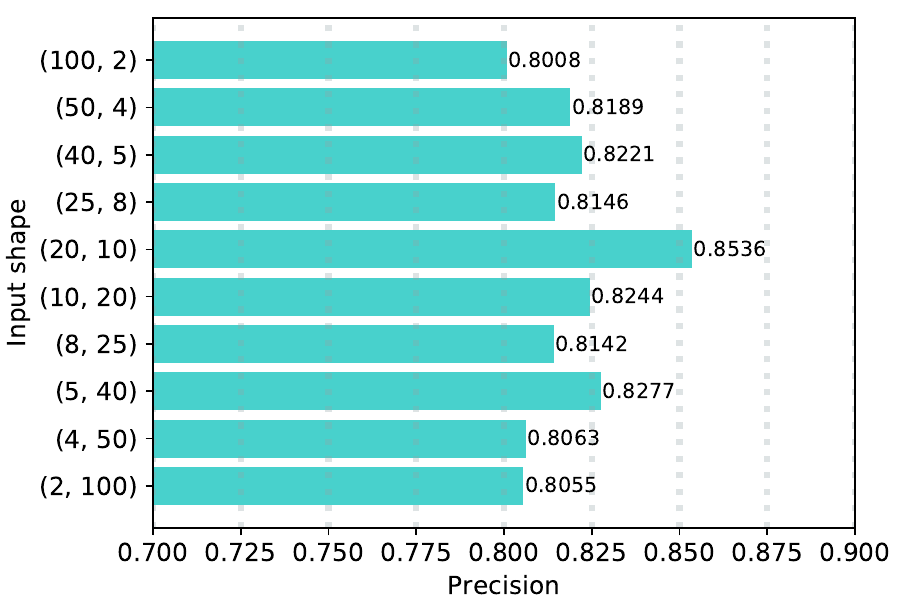}}
\subfigure[]{
\includegraphics[width=.48\linewidth]{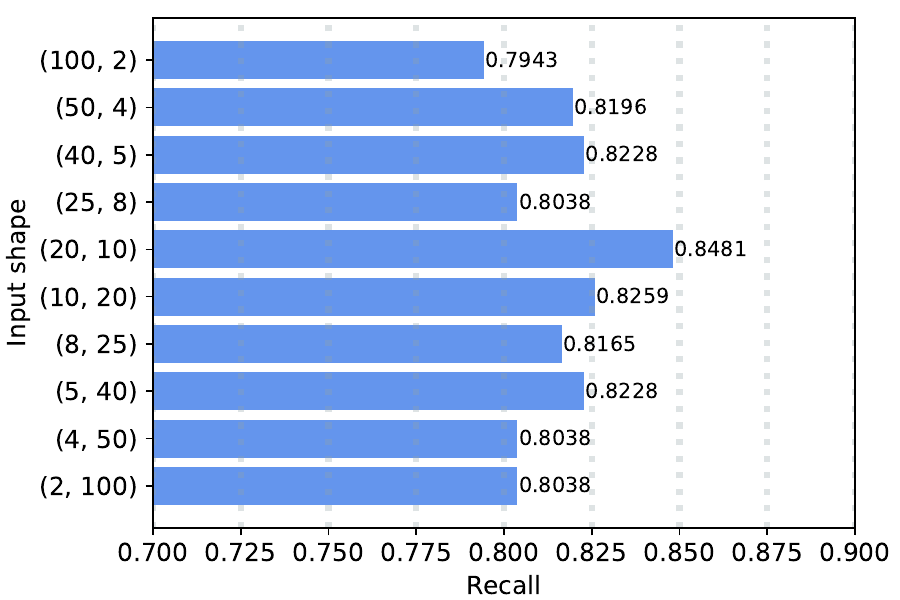}}
\vspace{-0.15in}
\caption{The effect of input shape in MDRR on the marine animal dataset.
}
\label{fig:auto}
\end{figure}

\vspace{-0.1in}
\subsubsection{Visualization of Clustering Results}
\label{sec:vis}
\begin{figure}[h]
\centering
\subfigure[t-SNE]{
\label{fig:tsne}
\includegraphics[width=.48\linewidth]{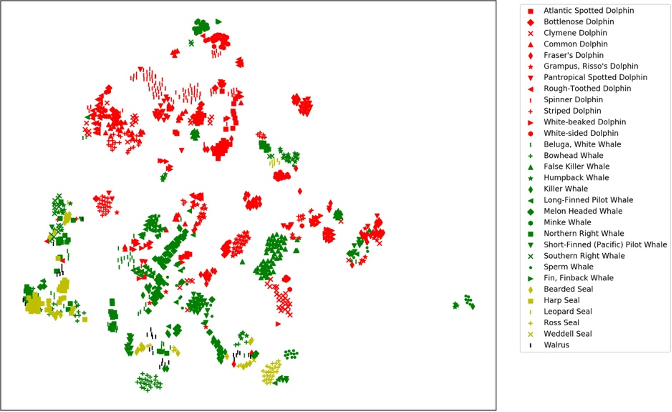}}
\subfigure[Hierarchical clustering]{
\label{fig:hier}
\includegraphics[width=.48\linewidth]{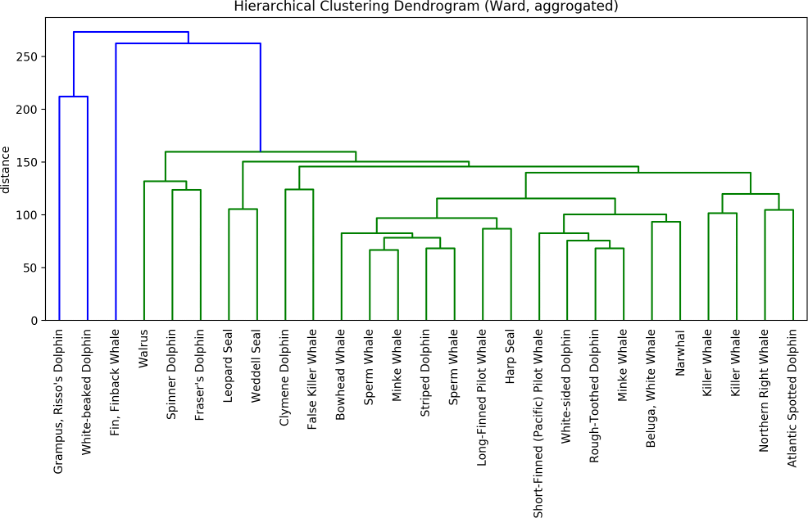}}
\vspace{-0.15in}
\caption{The learned representations from our model on the marine animal dataset. 
}
\label{fig:vis}
\end{figure}

We visualize clustering results from the perspective of t-distributed Stochastic Neighbor Embedding (t-SNE) \cite{cieslak2020t} and hierarchical clustering \cite{nielsen2016hierarchical} using the learned representations by our model in Fig. \ref{fig:tsne} and \ref{fig:hier}. 
In the t-SNE results, similar species within these marine groups tended to cluster closely together, reflecting their similar vocalization audio frequencies and resulting in overlapping areas in Fig. \ref{fig:tsne}. For example, Minke Whales, Humpback Whales, and Killer Whales form a closely-knit cluster.  This demonstrates that our proposed model effectively captures relevant features from audio data. However, some closely related species exhibited larger distances due to environmental noise, such as ship engines, which can mask their signals, such as Short-Finned (Pacific) Pilot Whales and Fraser's Dolphins.

In the hierarchical clustering analysis shown in Fig. \ref{fig:hier}, closely related marine animals were generally grouped into the same clusters. For instance, the Sperm Whale and Long-Finned Pilot Whale appear very close, while the Leopard Seal and Weddell Seal are clustered together. However, it's worth noting that different species were also grouped together in some cases, such as the Northern Right Whale and Atlantic Spotted Dolphin. This can be attributed: to 1) These marine animals may share similar audio frequencies; and 2) environmental noise, such as wind and wave sounds, might have influenced the clustering results to some extent. 

\vspace{-0.25in}
\section{Conclusion}
We propose a robust approach for automatic animal sound classification, addressing challenges in variable sound lengths, diverse frequencies, and noisy data. We introduce innovative techniques like MFCC feature matrix rearrangement, data reduction using autoencoders, and a deep learning model with Bi-LSTM and attention mechanisms. We contribute an animal sound benchmark dataset composed of marine animals and birds. Experiments on real-world datasets reveal significant performance improvements.

\bibliographystyle{IEEEbib}
\small
\bibliography{refs}

\begin{thebibliography}{10}

\bibitem{gasc2015acoustic}
Amandine Gasc, Sandrine Pavoine, Laurent Lellouch, Phillipe Grandcolas, and J{\'e}r{\^o}me Sueur,
\newblock ``Acoustic indices for biodiversity assessments: Analyses of bias based on simulated bird assemblages and recommendations for field surveys,''
\newblock {\em Biological Conservation}, vol. 191, pp. 306--312, 2015.

\bibitem{qi2008soundscape}
Jiaguo Qi, SH~Gage, Wooyeong Joo, Brian Napoletano, and S~Biswas,
\newblock ``Soundscape characteristics of an environment: A new ecological indicator of ecosystem health,''
\newblock {\em Wetland and water resource modeling and assessment}, pp. 201--211, 2008.

\bibitem{tucker2014linking}
David Tucker, Stuart~H Gage, Ian Williamson, and Susan Fuller,
\newblock ``Linking ecological condition and the soundscape in fragmented australian forests,''
\newblock {\em Landscape Ecology}, vol. 29, pp. 745--758, 2014.

\bibitem{phyu2009survey}
Thair~Nu Phyu,
\newblock ``Survey of classification techniques in data mining,''
\newblock in {\em IMECS}, 2009, vol.~1, pp. 727--731.

\bibitem{fu2010learning}
Zhouyu Fu, Guojun Lu, Kai~Ming Ting, and Dengsheng Zhang,
\newblock ``Learning naive bayes classifiers for music classification and retrieval,''
\newblock in {\em ICPR}, 2010, pp. 4589--4592.

\bibitem{shen2011automatic}
Peipei Shen, Zhou Changjun, and Xiong Chen,
\newblock ``Automatic speech emotion recognition using support vector machine,''
\newblock in {\em EMEIT}, 2011, vol.~2, pp. 621--625.

\bibitem{wei2020research}
Pengcheng Wei, Fangcheng He, Li~Li, and Jing Li,
\newblock ``Research on sound classification based on svm,''
\newblock {\em Neural Computing and Applications}, vol. 32, pp. 1593--1607, 2020.

\bibitem{mehyadin2021birds}
Aska~E Mehyadin, Adnan~Mohsin Abdulazeez, Dathar~Abas Hasan, and Jwan~N Saeed,
\newblock ``Birds sound classification based on machine learning algorithms,''
\newblock {\em Asian Journal of RCS}, vol. 9, no. 4, pp. 1--11, 2021.

\bibitem{vimal2021mfcc}
B~Vimal, Muthyam Surya, VS~Sridhar, Asha Ashok, et~al.,
\newblock ``Mfcc based audio classification using machine learning,''
\newblock in {\em ICCCNT}, 2021, pp. 1--4.

\bibitem{stowell2016bird}
Dan Stowell, Mike Wood, Yannis Stylianou, and Herv{\'e} Glotin,
\newblock ``Bird detection in audio: a survey and a challenge,''
\newblock in {\em MLSP}, 2016, pp. 1--6.

\bibitem{rajanna2015deep}
Arjun~Raj Rajanna, Kamelia Aryafar, Ali Shokoufandeh, and Raymond Ptucha,
\newblock ``Deep neural networks: A case study for music genre classification,''
\newblock in {\em ICMLA}, 2015, pp. 655--660.

\bibitem{dumpala2017improved}
Sri~Harsha Dumpala and Sunil~Kumar Kopparapu,
\newblock ``Improved speaker recognition system for stressed speech using deep neural networks,''
\newblock in {\em IJCNN}, 2017, pp. 1257--1264.

\bibitem{fang2022fast}
Zheng Fang, Bo~Yin, Zehua Du, and Xianqing Huang,
\newblock ``Fast environmental sound classification based on resource adaptive convolutional neural network,''
\newblock {\em Scientific Reports}, vol. 12, no. 1, pp. 6599, 2022.

\bibitem{xie2022sliding}
Jie Xie and Mingying Zhu,
\newblock ``Sliding-window based scale-frequency map for bird sound classification using 2d-and 3d-cnn,''
\newblock {\em Expert Systems with Applications}, vol. 207, pp. 118054, 2022.

\bibitem{tsalera2022cnn}
Eleni Tsalera, Andreas Papadakis, Maria Samarakou, and Ioannis Voyiatzis,
\newblock ``Cnn-based segmentation and classification of sound streams under realistic conditions,''
\newblock in {\em Pan-Hellenic}, 2022, pp. 373--378.

\bibitem{fu2010survey}
Zhouyu Fu, Guojun Lu, Kai~Ming Ting, and Dengsheng Zhang,
\newblock ``A survey of audio-based music classification and annotation,''
\newblock {\em Multimedia}, vol. 13, no. 2, pp. 303--319, 2010.

\bibitem{paraskevas2003audio}
Ioannis Paraskevas and Edward Chilton,
\newblock ``Audio classification using acoustic images for retrieval from multimedia databases,''
\newblock in {\em EURASIP}, 2003, vol.~1, pp. 187--192.

\bibitem{amiriparian2017snore}
Shahin Amiriparian, Maurice Gerczuk, Sandra Ottl, Nicholas Cummins, Michael Freitag, Sergey Pugachevskiy, Alice Baird, and Bj{\"o}rn Schuller,
\newblock ``Snore sound classification using image-based deep spectrum features,''
\newblock 2017.

\bibitem{ren2018learning}
Zhao Ren, Nicholas Cummins, Vedhas Pandit, Jing Han, Kun Qian, and Bj{\"o}rn Schuller,
\newblock ``Learning image-based representations for heart sound classification,''
\newblock in {\em ICDH}, 2018, pp. 143--147.

\bibitem{an2022development}
Ji-Hee An, Na-Kyoung Koo, Ju-Hye Son, Hye-Min Joo, and Seungdo Jeong,
\newblock ``Development on deaf support application based on daily sound classification using image-based deep learning,''
\newblock {\em JOIV}, vol. 6, no. 1-2, pp. 250--255, 2022.

\bibitem{cohen2020challenges}
Madison Cohen-McFarlane, Rafik Goubran, and Bruce Wallace,
\newblock ``Challenges with audio classification using image based approaches for health measurement applications,''
\newblock in {\em MeMeA}, 2020, pp. 1--5.

\bibitem{foote1997content}
Jonathan~T Foote,
\newblock ``Content-based retrieval of music and audio,''
\newblock in {\em Multimedia storage and archiving systems II}, 1997, vol. 3229, pp. 138--147.

\bibitem{sahidullah2012design}
Md~Sahidullah and Goutam Saha,
\newblock ``Design, analysis and experimental evaluation of block based transformation in mfcc computation for speaker recognition,''
\newblock {\em Speech communication}, vol. 54, no. 4, pp. 543--565, 2012.

\bibitem{church2017word2vec}
Kenneth~Ward Church,
\newblock ``Word2vec,''
\newblock {\em Natural Language Engineering}, vol. 23, no. 1, pp. 155--162, 2017.

\bibitem{bouzalmat2014comparative}
Anissa Bouzalmat, Jamal Kharroubi, Arsalane Zarghili, et~al.,
\newblock ``Comparative study of pca, ica, lda using svm classifier,''
\newblock {\em Journal of ETWI}, vol. 6, no. 1, pp. 64--68, 2014.

\bibitem{zhao2009comparative}
Zhiqiang Zhao, Yi~Zhang, Jianming Hu, and Li~Li,
\newblock ``Comparative study of pca and ica based traffic flow compression,''
\newblock {\em Journal of HTRD}, vol. 4, no. 1, pp. 98--102, 2009.

\bibitem{delac2005independent}
Kresimir Delac, Mislav Grgic, and Sonja Grgic,
\newblock ``Independent comparative study of pca, ica, and lda on the feret data set,''
\newblock {\em Journal of IST}, vol. 15, no. 5, pp. 252--260, 2005.

\bibitem{cieslak2020t}
Matthew~C Cieslak, Ann~M Castelfranco, Vittoria Roncalli, Petra~H Lenz, and Daniel~K Hartline,
\newblock ``t-distributed stochastic neighbor embedding (t-sne): A tool for eco-physiological transcriptomic analysis,''
\newblock {\em Marine genomics}, vol. 51, pp. 100723, 2020.

\bibitem{nielsen2016hierarchical}
Frank Nielsen and Frank Nielsen,
\newblock ``Hierarchical clustering,''
\newblock {\em Introduction to HPC with MPI for Data Science}, pp. 195--211, 2016.

\end{thebibliography}

\end{document}